# Reduced Complexity Simulation of Wireless Sensor Networks for Application Development

**Gursel Serpen and Zhenning Gao**
Electrical Engineering and Computer Science
University of Toledo, Toledo, OH 43606 USA

**Abstract**

This paper presents an approach for low-cost simulation modeling for application development for wireless sensor networks. Computational complexity of simulating wireless sensor networks can be very high and as such must be carefully managed. Application-level code prototyping with reasonable accuracy and fidelity can be accomplished through simulation that models only the effects of the wireless and distributed computations which materialize mainly as delay and drop for the messages being exchanged among the motes. This approach employs the abstraction that all physical or communication and protocol level operations can be represented in terms of their effects as message delay and drop at the application level for a wireless sensor network. This study proposes that idea of empirical modeling of delay and drop and employing those models to affect the reception times of wirelessly communicated messages. It further proposes the delay and drop to be modeled as random variables with probability distributions empirically approximated based on the data reported in the literature. The proposed approach is demonstrated through development of a neural network application with neurons distributed across the motes of a wireless sensor network. Delay and drop are incorporated into wireless communications, which carry neuron output values among motes. A set of classification data sets from the Machine Learning Repository are employed to demonstrate the performance of the proposed system in a comparative context with the similar studies in the literature. Results and findings indicate that the proposed approach of abstracting wireless sensor network operation in terms of message delay and drop at the application level is feasible to facilitate development of applications with competitive performance profiles while minimizing the spatio-temporal cost of simulation.

# 1. Introduction

There is growing potential and interest in utilizing wireless sensor networks (WSNs) for a wide variety of applications in industry, agriculture, forestry, nature, ecology, environment to name a few. WSNs offer promise for being able to observe, monitor and control events, processes and dynamics which were simply too difficult to do so for a variety of reasons including the risk of danger and the lack of technology among many others. Developing and prototyping distributed software for wireless sensor network applications on actual hardware is very challenging due to cost and complexity. Therefore, as it is the case with many other domains, simulation offers a much easier and manageable option to substitute for the actual hardware for prototyping. New applications and protocols for WSNs can be implemented on simulators to verify the feasibility and to assess their performance. All the while it is also relevant to recognize that the simulation-based prototyping will only be able to approximate the real hardware-based development. Given the complexity, cost, and the time involved in setting up an entire physical test-bed, simulators still offer a worthwhile option in light of their apparent shortcomings of varying degrees for accuracy, completeness, resolution and fidelity.

Simulation of wireless sensor networks may incur very high computation cost depending on the size of the sensor network in terms of the mote count and the accuracy and fidelity of simulation such as emulation of the hardware, or simulation at bit-, packet- or application-level. Determining the appropriate level of simulation is of paramount importance to manage the computational complexity for establishing the feasibility of a simulation. Simulators for bit or packet level emulation or simulation are unnecessarily detailed for prototyping distributed application code. Therefore, it is desirable to develop simulation frameworks or tools for distributed applications to maintain the balance between good accuracy and reasonable computation cost or complexity. WSN simulators can be classified into three major categories based on the level of complexity: bit level, packet level and algorithm level. As the complexity goes up, the time and memory consumption of the simulation also grows. It is desirable to select the level of simulation based on the rigor requirements of the experiment. For instance, a timing-sensitive medium access control (MAC) protocol would probably need a bit-level simulation while an algorithm-level simulation is sufficient to test the prototype developed for an agriculture management application.

     As bit-level simulators model the CPU execution at the level of instructions or even cycles, they are often regarded as emulators. TOSSIM [1] is configurable as either bit- and packet-level simulator for WSNs that are made up of TinyOS-based motes. TOSSIM simulates the entire TinyOS execution by replacing hardware components with emulated equivalents. It uses the same code as is used on real motes. TOSSIM simulates protocols and applications developed using the nesC code, a dialect of C programming language, which also runs on actual hardware by mapping hardware interrupts to discrete events. TOSSIM can handle simulations of WSNs with up to a thousand motes [1]. Avrora [2] is another bit-level simulator that is open source and built using the Java programming language. It simulates a network of motes by running the actual microcontroller-specific code, and accurate

simulations of the devices and the radio communication. Avrora is capable of running a complete sensor network simulation with high timing accuracy, which incidentally results in high computational burden.

Packet-level simulators implement the data link and physical layers in a typical open system interconnection (OSI) network stack. The most widely used simulator is ns-2 [3], which is an object-oriented discrete event network simulator built in C++. ns-2 can simulate both wired and wireless networks. J-Sim[4] is another packet-level simulator that adopts loosely-coupled, component-based programming model, and supports real-time process-driven simulation. OPNET [5], on the other hand, is a commercial simulator, which provides a simulation environment with a rich set of standard modules, and is a good choice to simulate Zigbee-based networks.

Algorithm level simulators focus on the logic, data structure and presentation of algorithms. They do not consider detailed communication models, and they normally rely on some form of a graph data structure to illustrate the communication between motes. Shawn [6] is a simulator implemented in C++ that has its own application development model or framework based on so-called processors. The motes or nodes in Shawn simulator are containers of processors, which process incoming messages, run algorithms and emit messages. The design philosophy of Shawn is dictated by the following. There is no difference between a complete simulation of the physical environment (or lower-level networking protocols) and the alternative approach of simply using well-chosen random distributions on message delay and drop for algorithm design on a higher level, such as localization algorithms. From Shawn's point of view, the common simulators spend much processing time on producing results that are of no interest at all, thereby actually hindering productive research on the algorithm. The framework of Shawn replaces low-level effects with abstract and exchangeable models. Shawn simulates the effects caused by a phenomenon instead of the phenomenon itself. For example, instead of simulating a complete medium access control (MAC) layer including the radio propagation model, its effects (in terms of packet drop and delay) are modeled in Shawn. Consequently, the simulation time of a WSN through Shawn is significantly reduced compared to other simulators, and the choice of the implementation model is more flexible.

Shawn is a generic simulator to accommodate a comprehensive set of simulation cases for a large selection of applications, and as such, suffers to some degree from overhead associated with its generic nature. Much of this overhead can be eliminated if the philosophy of simulating the "effects rather than the phenomenon itself" is implemented directly into the algorithm of the application that needs to be simulated for distributed computation on a wireless sensor network. The proposed approach would be as follows. Any given application algorithm that is to be implemented through distributed computation would first be partitioned into subparts as in functional decomposition, and the communication interface among these subparts would be identified. Finally, the messaging requirements would be established and delay and likelihood of drop are injected into the communication pathways through which packets carry those messages. This approach employs the maximum "abstraction" while still modeling the essential aspects of wireless sensor network operation and communications. For the purposes of verification and validation of application-layer code, modeling of effects in terms of "delay and drop" is promising to deliver the appropriate level of accuracy and computational cost.

In the following sections, an empirical statistical model for the message delay and drop is introduced to facilitate the implementation of application-level prototyping with minimal computational and communications complexity. A wireless sensor network that serves as a distributed computing platform for training and testing a multi-layer perceptron neural network configured as a classifier is employed to demonstrate the utility of the proposed empirical delay and drop model for reduced complexity simulation.

## 2. Statistical Models for Message Delay and Drop

This section presents empirical modeling for delay and drop phenomena for message communications in wireless sensor networks [77]. Using the empirical data compiled from studies reported in the recent literature, probability distributions for message delay and drop as a random variables are defined.

*Modeling the Probability Distributions for Packet Drop and Delay Phenomena*

The probability of packet drop or delay during wireless transmission in WSNs is highly dependent on the specific implementation of the network and its protocol stack. There are many factors at play, such as topology of the network, routing and MAC protocols, network traffic load, etc. It is not desirable to have the model for the probability distribution for drop or delay limited to a certain scenario (using certain protocols, number of motes, or topology, etc.) since the results of such a study would not be applicable in general terms. The model to be developed instead should be generalized enough to be applicable for the widest variety of WSN realizations, implementations and applications possible. One readily available option to develop or formulate a model for packet delay and drop is to leverage the empirical data reported in the literature, which is the venue pursued in our study.

We conducted a survey to compile the empirical data for message delay and drop as reported in the literature [7-28]. The survey covered a comprehensive set of simulation scenarios and compiled a record of the simulation settings and results. The simulation settings included routing protocols, MAC protocols, simulator type, number of motes, field size, radio range and others such as traffic, source count, dead node count etc. The simulation information that was compiled included the delivery ratio and the delay, which were extracted from tables and figures in the surveyed literature.

*Modeling the drop probability distribution*

In order to build an empirical model for the drop probability distribution, a literature survey was performed to collect and compile simulation data for different WSN designs, with variations in the topology and the protocol stack, and applications. The packet delivery ratio that was recorded in each study is considered as the main variable. Denoting the packet delivery ratio as $p_{delivery}$, the probability of drop, $p_{drop}$, can be calculated as $p_{drop} = 1 - p_{delivery}$. Specific values for the packet delivery ratio versus mote count for a number of WSN topologies and protocol stack implementations were retrieved from the studies reported in [7-28].

The data points are chosen based on the following assumptions or observations:

1) The node count is one of the primary independent variables, which means the data is collected for different node count values.
2) The density of node distribution within the WSN topology will stay "approximately" the same although the node count may vary. This means that the area of deployment for the network or the transmission range should change to keep the node distribution density the same.

3) Other factors such as the changing network traffic load or the static or time-varying percentage of dead nodes could not be considered due to lack of sufficient empirical data, relevant simulations or experiments.

Establishing the above specifications is intended to ensure that packet drop probability is mainly, but not solely, affected by the number of transmission hops only, which is further employed to approximate the distance between the sender and receiver mote pair.

We first investigate the relationship between the probability of drop and mote count for a variety of routing protocols using the studies reported in literature [7-28]. The plots for different routing protocols for the probability of drop versus node count are shown in Figure 1. The routing protocols were QoS [16], Speed [16], GBR [7], LAR [18], LBAR [18], Opportunistic Flooding [14], AODVjr [18], BVR [10], DD [23], and EAR [7].

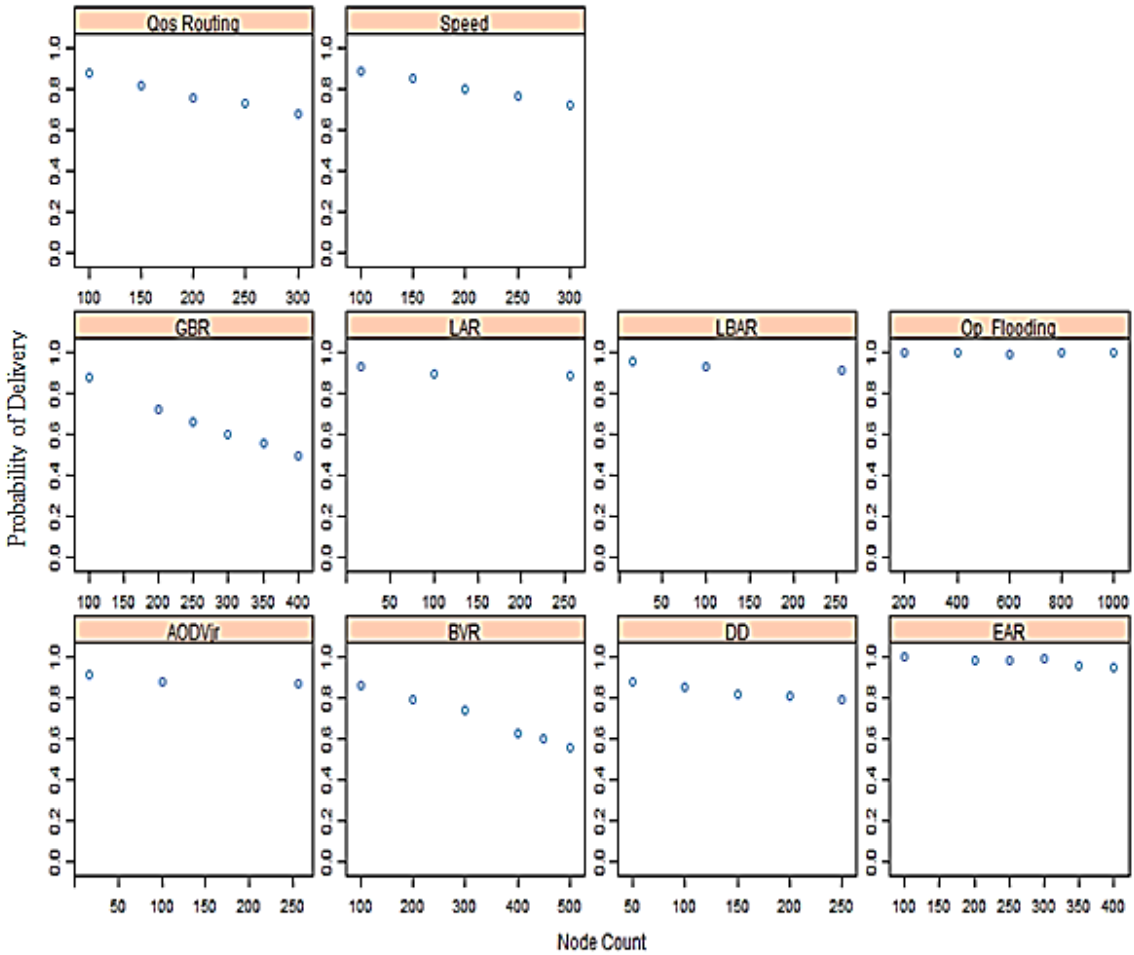

Figure 1: Plots for probability of delivery vs. mote count for various routing protocols

In Figure 1, the *x*-axis represents the node count, while the *y*-axis is the probability of delivery. Each individual plot is specific to a "routing protocol". Denoting the node count as $n_{nodes}$, Figure 1 shows that $p_{delivery}$ decreases when the value of $n_{nodes}$ increases. The relationship appears to be linear in general. Since these data are due to specific experiments, in order to generalize, it may not be a good idea to make the model fit the data precisely. Therefore, the linear regression (versus a polynomial) for fitting these data points is a reasonable option. Then the resultant empirical model is given by

$$p_{delivery} = \frac{(\beta_0 + \beta_1 \times n_{nodes})}{100},$$

where coefficients $\beta_0$ and $\beta_1$ are real numbers for the linear model and $n_{nodes} \in [0,1000]$. The probability of drop is calculated as

$$p_{drop} = 1 - \frac{(\beta_0 + \beta_1 \times n_{nodes})}{100}, \tag{1}$$

For the linear regression model for each plot shown in Figure 1, the coefficients $\beta_0$ and $\beta_1$ calculated for each case is shown in Table 1. In the case for the Opportunistic Flooding routing protocol, a special scenario arises: it can guarantee a successful delivery, which means the probability of drop is zero.

| Routing Protocol | $\beta_0$ | $\beta_1$ | **Empirical Model** |
|---|---|---|---|
| EAR | 100.82 | -0.0107 | $p_{drop} = (-0.82 + 0.043 \times n_{hops}^2)/100$ |
| GBR | 94.50 | -0.1130 | $p_{drop} = (5.5 + 0.45 \times n_{hops}^2)/100$ |
| BVR | 94.44 | -0.0760 | $p_{drop} = (5.6 + 0.076 \times n_{hops}^2)/100$ |
| QoS | 97.00 | -0.0980 | $p_{drop} = (3 + 0.049 \times n_{hops}^2)/100$ |
| Speed | 97.40 | -0.0840 | $p_{drop} = (2.6 + 0.042 \times n_{hops}^2)/100$ |
| LBAR | 95.79 | -0.0198 | $p_{drop} = (4.21 + 0.020 \times n_{hops}^2)/100$ |
| LAR | 92.57 | -0.0154 | $p_{drop} = (7.43 + 0.015 \times n_{hops}^2)/100$ |
| AODVjr | 90.57 | -0.0154 | $p_{drop} = (9.43 + 0.015 \times n_{hops}^2)/100$ |
| DD | 89.60 | -0.0440 | $p_{drop} = (10.4 + 0.088 \times n_{hops}^2)/100$ |
| Opportunistic Flooding | 100.00 | 0.0000 | $p_{drop} = 0$ |

Table 1: Coefficients $\beta_0$ and $\beta_1$ of the linear model for different routing protocols reported in Figure 1

It is reasonable to assume that the number of hops can be used as the primary factor affecting the probability of drop for a message packet. Given a two-dimensional deployment topology for a WSN, let $n_{hops}$ denote the hop count between a source and a destination mote pair. Defining the $p_{drop}$ in terms of $n_{hops}$ is of interest. Given that we know the value for $n_{nodes}$, which is the number of motes in the WSN, a relationship between $n_{nodes}$ and $n_{hops}$ needs to be derived for a given specific deployment topology. For instance, as shown in Figure 2, consider that $n_{nodes}$ motes are uniformly randomly distributed in a square deployment area. Consequently, the number of motes along any edges will be approximately $\sqrt{n_{nodes}}$. Assume the source mote is located at the center, while the sink node is close to one of the corners. Average hop count for a message, $n_{hops}$, can be approximated by the length of the diagonal in terms of number of hops divided by two. Then the relationship of $n_{nodes}$ and $n_{hops}$ is given by $n_{nodes} = 2n_{hops}^2$.

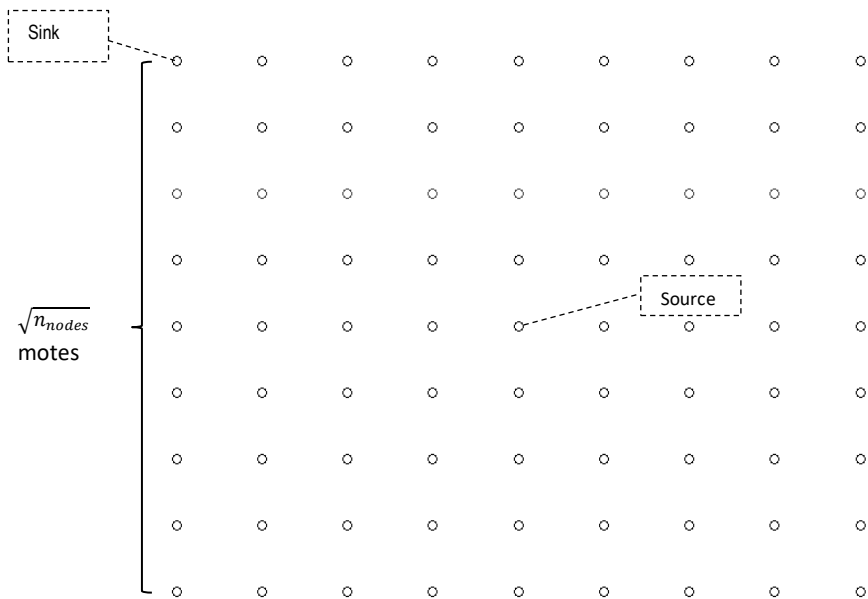

Figure 2. Illustration of relationship between $n_{hops}$ and $n_{nodes}$

In the worst case for a source and sink mote pair where the motes are located at the end points of a given diagonal, the relationship between these two variables becomes $n_{nodes} = n_{hops}^2/2$. Consequently, the hop count values are bounded as follows: $1 \leq n_{hops} \leq \sqrt{2 n_{nodes}}$. The square topology assumed for the above analysis is a reasonable approximation to many of the deployment realizations. If necessary, other topologies can also be readily analyzed following a similar approach. In more general terms, the relationship between $n_{nodes}$ and $n_{hops}$ can be represented as

$$n_{nodes} = \tau \times n_{hops}^2, \qquad (2)$$

where τ is the coefficient whose value is positive and will vary based on a number of WSN-related parameter settings including the shape of the topology and the density of mote deployment. As an example, consider the QoS-based routing protocol implementation in the study reported in [16]. The topology is a unit square and the two sink nodes are placed in the lower corners of a square deployment area. The nodes in the upper right report to the sink in the bottom left and the nodes in the upper left report to the sink in the bottom right. The distance between the source and the sink nodes is approximately the diagonal of the square area. Accordingly, the coefficient τ has a value of $\sqrt{2}$.

In the linear regression curves obtained for each routing protocol earlier, the coefficients $\beta_0$ and $\beta_1$ and the parameter $n_{nodes}$ values are substituted to yield the empirical models shown in Table 1. The empirical models in Table 1 indicate that the models for GBR and Opportunistic Flooding routing are exceptions as their performances vary significantly when compared to the performances of the other cases. The GBR protocol would not be appropriate for a large network, and the Opportunistic Flooding would incur too much delay to guarantee the delivery.

Page **12** of **49**

Therefore, the models for GBR and Opportunistic Flooding are not considered any further. We consider all the remaining models in Table 1 to set a range for $\beta_0$ and $\beta_1$ parameters in Equation 3. Accordingly, the empirical model for $P_{delivery}$ is defined in general terms as

$$p_{delivery} = \left(\beta_0 + \beta_1 \times \tau \times (n_{hops})^2\right)/100, \tag{3}$$

where the range of values for $\beta_0$ is (-1, 11), and the range of values for $\beta_1$ is (0.013, 0.09). When this model is employed in the simulation study $\beta_0$ and $\beta_1$ values can be generated based on how "similar" the routing protocol to one of the given in Table 1 or are generated randomly using a uniform distribution in their corresponding ranges if no strong similarity can be established.

*Modeling the delay probability distribution*

The literature survey indicated that the length of message delay varies from 10 ms to 3000 ms for different implementations (such as variations in number of nodes, routing protocol, MAC protocol, packet length, traffic load etc.) [7-28]. Figure 3 shows the histogram for the delay based on the survey data.

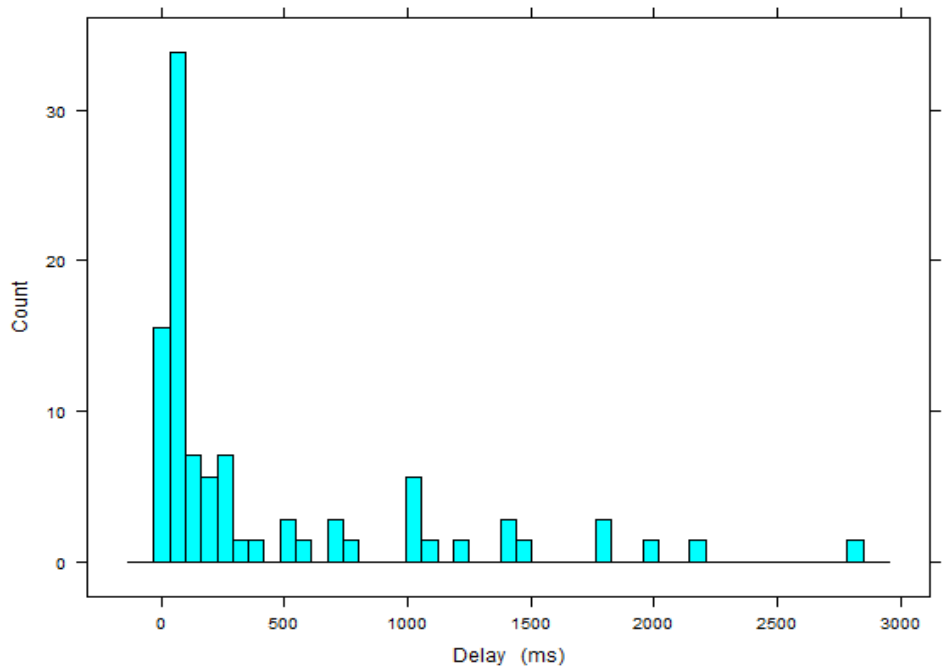

Figure 3. Histogram of Delay for Empirical Data Reported in Literature

In a design scenario where one or more neurons are embedded into a given mote, the exchange of neuron outputs among motes will be subject to certain delay that is inherent in wireless communications. This delay, which will dictate the duration of a waiting period by a given neuron for its inputs to arrive from other neurons on other motes is not a fixed value but rather a random variable. This delay-induced wait time will be denoted as $t_{wait}$. Note that $t_{wait}$ is both application dependent and network dependent: as indicated in Figure 3, its value was found to vary from 10 ms to 3000 ms per the literature survey. For a specific network, the delay between for a pair of motes from one pair to another varies substantially, and even for the same pair of motes the delay variance is significant. Additionally, the maximum delay could be much larger than the mean delay [29,35]. In simulating a neural network embedded across motes of a wireless sensor network, the $t_{wait}$ is set according to the mean delay value and the specific network topology to make sure that a good number of inputs successfully arrive for any given neuron so that it can calculate its own output.

Per the literature, a specific delay distribution is highly dependent on many factors such as the MAC protocol, traffic, queue capacity, channel quality, back-off time setting in MAC protocol, etc. [29-31,33 and 35]. Therefore, it is impossible to get a highly accurate model of delay distribution considering that so many factors play a role in affecting its value. A reasonably good but approximate model however can be formulated by using the Gaussian distribution. It has been used by others to model the delay distribution [30-34]. Empirical evidence suggests that the delay distribution is truncated and heavy-tailed [33,35]. Consequently, we consider a truncated Gaussian distribution for modeling the delay variance. The truncated Gaussian distribution is for a normally distributed random variable whose value is bounded. Suppose a random variable $x$ has a normal distribution with $N(\mu, \sigma^2)$ and lies within a range of ($a$, $b$), then $x$ conditional on $a<x<b$ has a truncated normal distribution. Its probability density function $f$ is given by

$$f(x; \mu, \sigma, a, b) = \frac{\frac{1}{\sigma}\varphi(\frac{x-\mu}{\sigma})}{\Phi(\frac{b-\mu}{\sigma}) - \Phi(\frac{a-\mu}{\sigma})}, \qquad (4)$$

where $x$ is the random variable; $\mu$ represents the mean; $\sigma$ represents the standard deviation; $a$ represents the minimum value; $b$ represents the maximum value; $\varphi(\cdot)$ is the probability density function of the standard normal distribution; and $\Phi(\cdot)$ is the cumulative distribution function.

The "delay" for a given neuron output is positive integer valued and quantified to correlate with the number of pattern presentations. This parameter value is computed by summing per hop delays for the total number of hops between the sending-receiving neuron pair and dividing this sum by value of the $t_{wait}$ parameter. Truncated Gaussian distribution is used as the model for the "per hop delay" parameter. In other words, the computation employs the following steps:

$$Delay = floor\left(\frac{\sum_1^{n_{hops}} per\ hop\ delay}{t_{wait}}\right) \qquad (5)$$

where the *floor*() function truncates the digits after the decimal point to extract the integer part of a given real number argument. The parameter $t_{wait}$ is defined in terms of three other parameters as follows:

$$t_{wait} = \vartheta \times \mu \times l_{max}, \qquad (6)$$

where $\vartheta$ is a coefficient to be set in the simulator, $\mu$ is the mean value of the truncated Gaussian distribution, and $l_{max}$ is the max hop count of the topology being considered. For ease of computation, the mean value of truncated Gaussian distribution will be normalized to relocate it to the value of 1.0. Based the empirical studies in the literature [29,30], other parameters of the truncated Gaussian distribution are set as *a*=0.3, *b*=5 and σ=0.6.

# III. WSN-MLP Design

This section will present the design of a hybrid system named as WSN-MLP, based on a wireless sensor network (WSN) and a multilayer perceptron (MLP) feedforward neural network (NN) for classification tasks. The WSN is configured as a parallel and distributed processing platform for training of and classification by an MLP NN.

**Multilayer Perceptron Neural Network**

An MLP is a layered artificial neural network (ANN) architecture as shown in Figure 4, where neurons are assigned to different layers with distinct functional roles. An MLP consists of at least three types of layers; namely an input layer, one or more hidden layer(s), and an output layer. The input layer is not considered a "true" layer because it does not perform any sort of computation. Its role is to receive domain-specific external inputs, and distribute those to neurons in the hidden layer. An MLP may have one or more hidden layers, which receive inputs from preceding layers (input or previous hidden layers). Outputs from a hidden layer are supplied as inputs to neurons either in the subsequent hidden layer (if there is one) or the output layer. Each neuron in a hidden layer individually performs processing of its inputs. The associated computation typically entails scalar product of its input and weight vectors and processing the resulting scalar value through a nonlinear mapping or activation function, which is usually sigmoidal shaped [36]. The output layer presents the result of the computation collectively performed by the network to the external world.

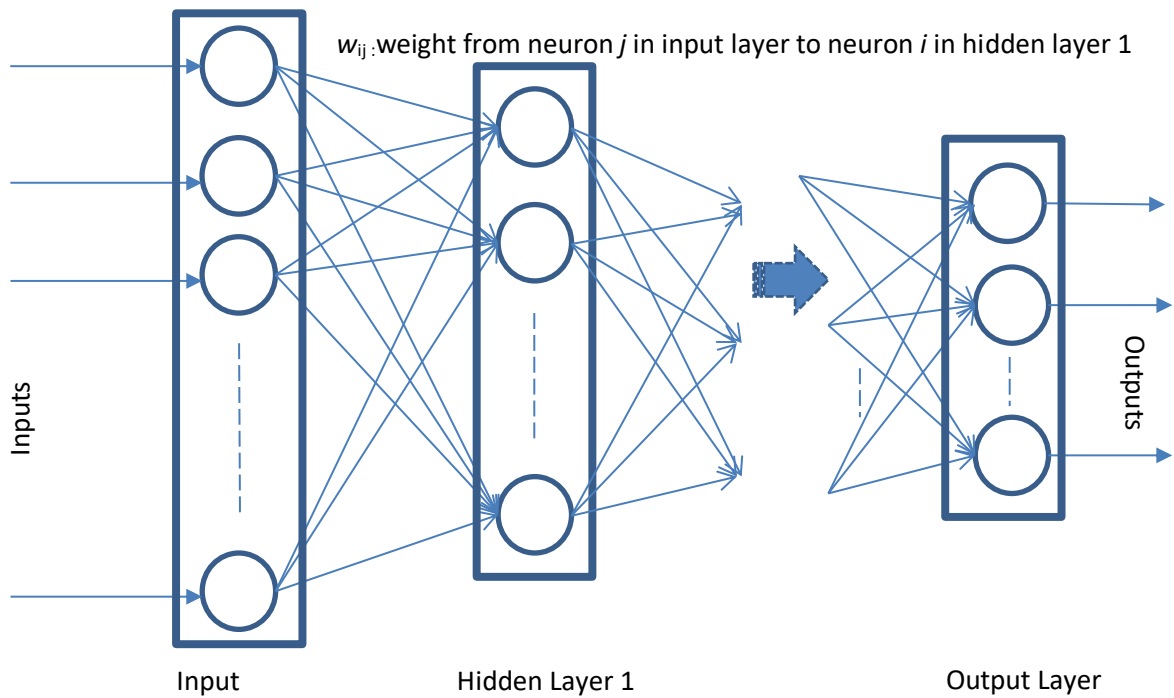

Figure 4. MLP neural network architecture

The network exhibits a high degree of connectivity, meaning all neurons in one layer are connected to all neurons in the next layer [36]. MLP is a feed-forward neural network, which means the information moves in only one direction, namely forward, from the input neurons through the hidden neurons, and to the output neurons.

In its most basic form, an MLP neural network utilizes a supervised learning technique called back-propagation for training. The training consists of two phases: a forward phase and a backward phase. During the forward phase, the in-network processing or computation takes place, where input data are propagated forward through the MLP neural network from the input layer to the output layer while the weights remain unaltered. At the start of the backward phase, an error vector for output layer neurons is calculated based on the difference between the desired and the computed output values. Then the error vector is propagated back to the hidden layer(s).

The error vector for hidden layer(s) is next calculated. The weights are updated based on the values of the error and the input pattern vectors. The back-propagation learning algorithm performs a gradient descent search in the weight space for the lowest value of the error function.

There are several variations for gradient descent based back-propagation training algorithm. It is a well-established fact in the literature that the basic back-propagation algorithm can be very slow for training even for a simple MLP neural network topology [37,38]. In the context of a wireless sensor network implementation, what is needed is a fast learning algorithm for the MLP neural network, which projects minimal computational cost; requires minimal centralized coordination if any; is realizable through incremental learning (vs. batch learning); and can be implemented in a parallel and distributed manner. The steepest descent is particularly slow when there is a long and narrow valley in the error function surface [39]. This problem can be addressed for the most part using the so-called momentum term, which allows a network to respond not only to the local gradient, but also to recent trends in the error surface. The momentum term helps average out the oscillations along the short axis while at the same time adds up contributions along the long axis [40]. It is well known that such a term greatly improves the speed of learning. Momentum is incorporated into back propagation learning by making weight changes equal to the sum of a fraction of the most recent weight change and the new change suggested by the gradient descent rule. When using the back propagation with momentum, the k-th update value for a weight $w_{ij}$ is given by

$$\Delta w_{ij}(k) = -\gamma g_{ij} + \alpha \Delta w_{ij}(k-1) \qquad (7)$$

where $\gamma$ is the learning rate, $\alpha$ is the momentum rate, $g_{ij}$ is gradient of the error with respect to the weight vector, and $\Delta w_{ij}(k-1)$ is the former update value for the $w_{ij}$. During training of each neuron, all that is needed is to store the previous local weight update values. This makes the gradient descent with momentum algorithm feasible to implement in a distributed manner for the MLP that is embedded within a WSN.

**Embedding MLP NN within WSN**

Consider a multilayer perceptron (MLP) type artificial neural network (ANN) with at least three layers of neurons, namely an input layer, one (or more) hidden layer(s), and an output layer. We consider the case where each participating mote houses exactly one neuron for maximizing the parallel and distributed nature of computations: if the number of motes is greater than the number of neurons, some of the motes will not participate in the neurocomputing effort.

Output of a neuron embedded within a mote will need to be communicated to neurons on other motes through wireless communication channels or over the air for a wireless sensor network realization. Packets are subject to delay and drop during wireless transmission due to medium access (such as channel being busy or collision of packets), outgoing or incoming message processing, multi-hop communications, and routing algorithms among many other factors. Meaningful simulation of WSN communications for an MLP NN training requires that such delay and drop be modeled as accurately as reasonably possible while maintaining low computational cost.

Outputs of neurons in one layer must be communicated to inputs of neurons in the next layer during training and following the deployment. Since the wireless communication of such packets that carry neuron output values is accomplished typically through multi-hop routing, it is reasonable to assert that the delay due to medium access, packet processing, and the hop-count among others will be mainly affected by the distance (or the equivalently the number of hops) between the sending and receiving neurons (or motes). Although it may may be affected by the actual routing protocol chosen, the distance of the routing path for a packet can be approximated by the hop count, which is measurable through various approaches [7]. The likelihood of packet drop carrying a neuron output increases as the number of hops increases between the sender-receiver neuron (or the corresponding mote) pair. Accordingly, the hop count may be employed as the primary factor affecting the amount of delay and the likelihood of drop for packets carrying neuron outputs. When delay occurs and its value varies and, in the worst case, the drop happens, a procedure needs to be developed to make available past values of the output for the neuron whose output is delayed or dropped.

*Modeling drop and delay for packets carrying neuron outputs*

For each wireless packet carrying neuron output values, computation of values for the delay and drop random variables needs to be accomplished. The first step entails the computation of the distance between a pair of motes (neurons). For ease of computation and without losing generality, we assume that motes are randomly deployed across a two-dimensional field. The output-layer neurons are placed in the smaller square field in the middle of the larger square field of hidden-layer nodes as shown in Figure 5. We further assume that the number of neurons in the hidden layer is much greater than those in the output layer as is typical in most, if not all, applications. This deployment arrangement facilitates the average of distances between node pairs to be minimized, and also keeps the variance of distance as small as possible.

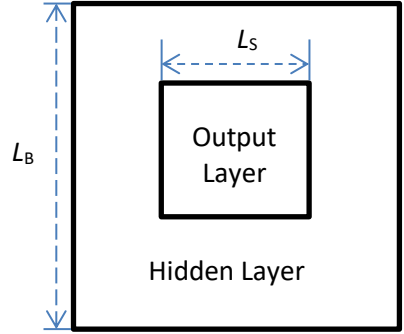

Figure 5. Placement of MLP NN layers within a two-dimensional square WSN topology

Let $N_{HL}$ and $N_{OL}$ represent the number of neurons (motes) for the hidden layer and the output layer, respectively. Corresponding to the case shown in Figure 5, the length of one side of the larger square is given by $L_B = \sqrt{N_{HL} + N_{OL}}$. The hidden layer neurons are deployed uniformly randomly within this square excluding the area occupied by the inner smaller square. The output layer neurons are deployed in the inner square whose side length (denoted as $L_S$) is given by $L_S = \sqrt{N_{OL}}$. For each neuron, its coordinates are generated randomly in the corresponding square area. The Euclidian distance between a pair of motes (neurons) $i$ and $j$ with coordinates $(x_i^{HL}, y_i^{HL})$ and $(x_j^{OL}, y_j^{OL})$ is given by $ED_{ij} = \sqrt{(x_i^{HL} - x_j^{OL})^2 + (y_i^{HL} - y_j^{OL})^2}$. Since the Euclidean metric value is in terms of the number of motes, this distance would also correspond roughly to the distance in terms of the number of (transmission) hops, namely $n_{\text{hops}}$.

*Delay model for wireless transmission of neuron outputs*

For each wireless communication link between a pair of motes, the probability of drop $p_{drop}$ for a message packet is calculated based on the hop count between them using Equation 2. The value of $t_{wait}$ is application specific and set by Equation 6. In our experiments, we change the value of the coefficient $\vartheta$ to observe the neural network performance under different conditions. Then the delay amount for each transmission is generated using Equation 5. Let the following definitions hold for the delay and drop model under the assumption that each mote is embedded with a single neuron:

$z$:        a uniformly-distributed random number instance in the range [0,1];

$i$:        index for sending mote;

$j$:        index for receiving mote;

$m_i, m_j$: labels for sending and receiving motes or neurons, respectively;

$d_{ij}$:       delay (positive integer-valued) for the packet sent by mote (neuron) $m_i$ to mote (neuron) $m_j$;

**s**$_{ij}$:     one-dimensional array of storage for communication between motes $m_i$ and $m_j$, where the array index corresponds to pattern presentation number (or the sequence number for the number of patterns presented in a single training epoch), and contents the hold presentation number of the most recently received pattern;

**r**$_j$:      one-dimensional array for neuron $j$ where each array element holds the pattern presentation number for the most recently generated output from the corresponding neuron (mote) that has arrived at mote $m_j$. More specifically, the $\mathbf{r}_j[i]$ holds the pattern presentation number for the most recently generated packet among the packets which arrived at mote $m_j$ from mote $m_i$;

$\mathbf{o}_j$: one-dimensional array where $\mathbf{o}_j[k]$ holds the value of output for neuron $m_j$ computed during the *k*-th training pattern presentation.

The pseudo-code in Figure 6 defines the implementation of delay and drop models for messages between motes $m_i$ and $m_j$. Mote $m_i$ transmits a packet carrying the output value of its neuron to mote $m_j$ after each pattern presentation unless there is delay. At a given pattern presentation iteration *k*, the first step is to update the array $\mathbf{s}_{ij}$. Initial decision is to determine if the packet should be dropped or not. The packet is considered as "not dropped" if the randomly generated number *z* is greater than the probability of drop ($p_{drop}$). If the packet is not dropped, then the delay amount $d_{ij}$ is generated. The presentation sequence number for the current pattern *k* is stored in the array $\mathbf{s}_{ij}$ at the position indexed by $k + d_{ij}$; this means the packet that carries the neuron output value and destined from mote $m_i$ to mote $m_j$ during pattern presentation *k* would arrive at mote $m_j$ during pattern presentation $k + d_{ij}$. If the random number *z* is not larger than the probability of drop, $p_{drop}$, then the packet is considered as "dropped." Then, $\mathbf{s}_{ij}$ array values won't be updated which means the packet from mote $m_i$ to mote $m_j$ during pattern presentation *k* won't arrive at all. The second step is to update $\mathbf{r}_j[i]$, which holds the pattern presentation sequence number for the most recently generated packet among the packets which arrived at neuron $m_j$ from $m_i$ (an example is given in the following paragraph). If $\mathbf{s}_{ij}[k]$ is larger than $\mathbf{r}_j[i]$, $\mathbf{r}_j[i]$ value is updated to that of the $\mathbf{s}_{ij}[k]$. This means that there is a packet that arrived during pattern presentation k, which is more recent compared to any other output, which arrived from that same neuron. The content of array element $\mathbf{o}_j[\mathbf{r}_j[i]]$ is used for updating neuron's output on mote $m_j$.

```
Step 0: a. Initialize the array s_ij[] to 0.
        b. Initialize r_j[i] to 0.
During kth pattern presentation, where k=1,2,3,
Step 1: If z > p_drop
                Generate the d_ij value using the delay model.
                Update the array s_ij[]: s_ij[k + d_ij] = k
Step 2: If s_ij[k] > r_j[i]
                Update r_j[i]: r_j[i] = s_ij[k].
```

Figure 6. Pseudo-code for implementation of delay and drop model.

An illustrated example demonstrating the use and the realization of the delay and drop modeling is presented Figure 7. The example shows the execution during pattern presentations 17 and 18. The original value of the arrays $\mathbf{s}_{ij}$ and $\mathbf{r}_j[i]$ are assumed as shown in the same figure. In the first step of presentation of pattern 17, assume $z$ is larger than $p_{drop}$, and the generated delay amount $d_{ij}$ is 1. This means the packet generated during this pattern presentation would arrive at the next pattern presentation, namely during the presentation of pattern with sequence number 18. The content of $\mathbf{s}_{ij}[18]$ is updated to 17. In the second step of 17[th] pattern presentation, the packet that arrived at this pattern presentation is from 15th pattern presentation (the content of $\mathbf{s}_{ij}[17]$ is 15). It is older than the packet that arrived during the 16[th] pattern presentation. Since 15 is less than 16 where the latter is stored in $\mathbf{r}_j[i]$, $\mathbf{r}_j[i]$ is not updated. The neuron output value (from $m_i$) stored in $\mathbf{o}_j[16]$ is used for updating the output or weights of the neuron on mote $m_j$. Next, consider the pattern presentation sequence 18. During the first step of $18^{th}$ pattern presentation, assume $z$ is larger than $p_{drop}$, and the delay amount $d_{ij}$ is determined to be 0. This means the packet arrived on time (it arrived during the current pattern presentation). The content of $\mathbf{s}_{ij}[18]$ is updated to 18 (18 replaces 17). In the second step of $18^{th}$ pattern presentation, the packet that arrived during this pattern presentation is from pattern presentation 18 (the content of $\mathbf{s}_{ij}[18]$ is 18). 18 is greater than the content of $\mathbf{r}_j[i]$ which is equal to 16, so $\mathbf{r}_j[i]$ is updated to be 18. The output value stored in $\mathbf{o}_j[18]$ is used for updating the neuron dynamics on mote $m_j$.

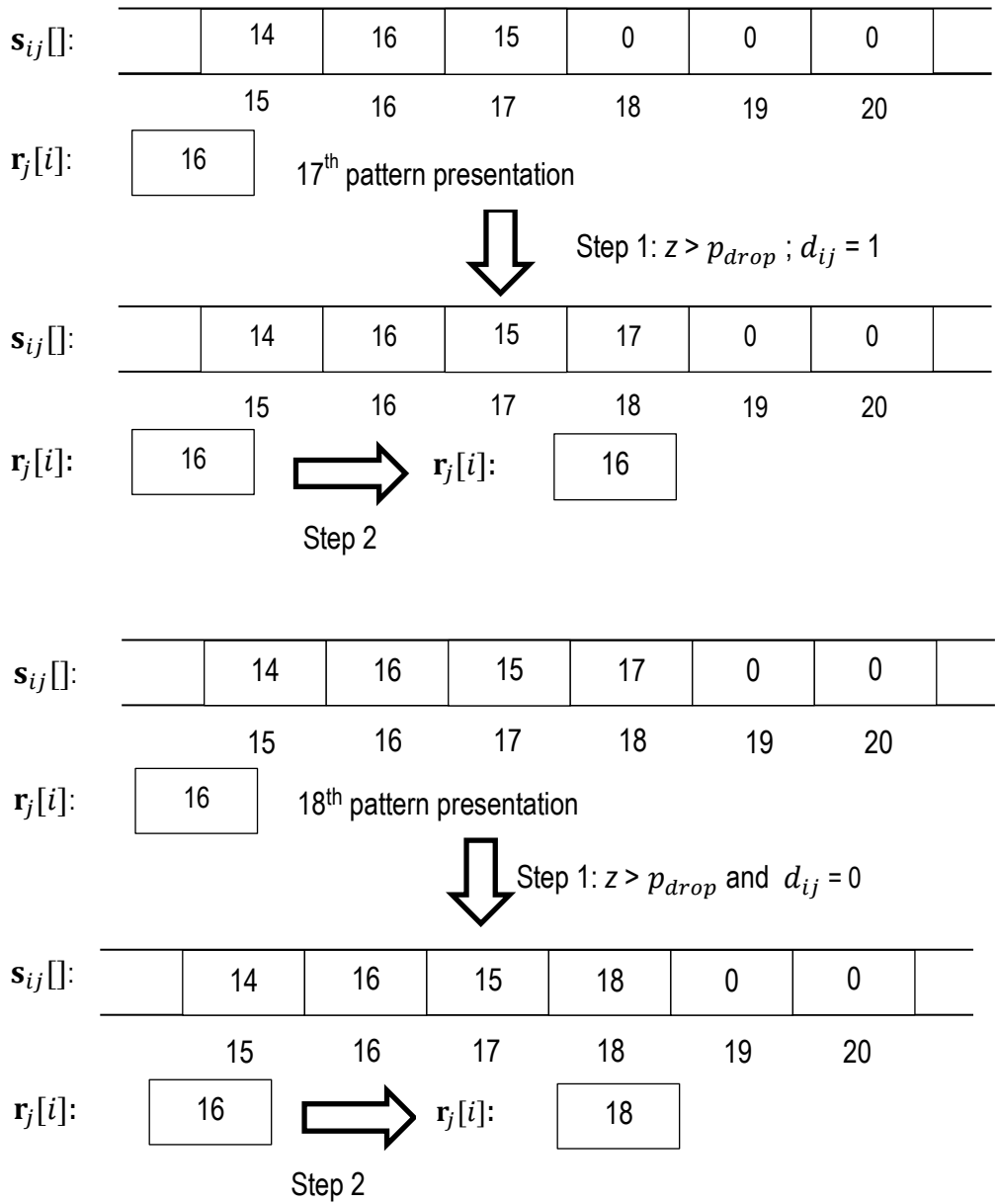

Figure 7. Example for the implementation of the delay and drop modeling

## IV. Simulation Study

This section starts with the description of the data sets employed in the simulation study and details the preprocessing implemented for utilization on the proposed wireless sensor network and multilayer perceptron (WSN-MLP) platform. Next, it discusses parameter settings for the MLP-BP algorithm including the number of hidden layer neurons, the rationale for training algorithm selection, and partitioning of the data set for training and testing among others.

**The Simulator**

The simulator was custom developed in-house and implemented in C++. It simulates the delay and drop effects on the transmission of neuron outputs as described earlier. It therefore provides a highly computationally-efficient simulation bypassing the details not relevant for performance assessment associated with application development for a wireless sensor network context.

After the simulator initialization phase, the training of the MLP neural network begins. Training of the MLP network entails forward propagation, back propagation, and weights update. After each complete iteration over the entire training dataset, the MLP performance is validated on the testing data. The effects due to delay and drop are modeled for the transmission of outputs from the hidden layer neurons in the forward propagation step, and the error signals generated at the output layer and communicated back to hidden layer neurons in the backward propagation step. The training related performance data includes the classification accuracy on the test data, mean squared error computed on the test data, number of training iterations, the confusion matrix on the test data, percentage of drop and delay of packets, and the converged values of the weight matrix between the hidden and output layers. Distribution of training patterns from the gateway mote to other motes is accomplished through single-hop transmission.

It is also assumed that potential delays or drop for the communications involving the gateway mode are not significant, and therefore can be ignored.

**Data Sets**

The simulation study is conceived to assess the scalability of the WSN-MLP design as the number of attributes, instances, and classes of a problem domain increase. Six data sets are used for the simulation study. They differ for their number of attributes, number of classes, number of instances, and the domain. As Equation 3 indicates, the effect of packet drop would be significant only when the mote count is large enough. Due to this reason, the simulation study employs data sets with attribute counts of up to 5000. Such a large attribute count also results in an MLP neural network that has hundreds of hidden layer neurons. The characteristics of datasets are presented in the Table 2. The data sets are from the UCI Machine Learning Repository (UCI) and discussed in the following subsections [41].

| Dataset | Attribute Count | Instance Count | Class Count | Class Distribution | Problem Domain |
|---|---|---|---|---|---|
| Iris | 4 | 150 | 3 | 1:1:1 | Life |
| Wine | 13 | 175 | 3 | 1:1.2:0.8 | Physical |
| Ionosphere | 34 | 351 | 2 | 1.3:0.7 | Physical |
| Dermatology | 34 | 358 | 6 | 1.9:1:1.2:0.8:0.8:0.3 | Life |
| Handwritten numerals | 240 | 2000 | 10 | 1:1:1:….1 | Word |
| Isolet | 617 | 7797 | 26 | 1:1:1:….1 | Speech |
| Gisette | 5000 | 7000 | 2 | 1:1 | Word |

Table 2. Characteristics of data sets employed in the simulation study

Iris is one of the best known data sets in the pattern recognition field. This data set contains 3 classes of 50 instances each, where each class refers to a type of Iris plant [41]. It contains 4 attributes as sepal length, sepal width, petal length, and petal width.

The wine data set is the result of a chemical analysis of wines grown in the same region in Italy but derived from three different cultivars [41] or classes. The 13 attributes relate to the quantities of 13 constituents found in each of the tree types of wines. Three classes (namely Class 1, Class 2, and Class 3) have 58, 71, and 48 instances, respectively.

The ionosphere data set contains radar data collected by a system in Goose Bay, Labrador [41]. This system consists of a phased array of 16 high-frequency antennas with a total transmitted power on the order of 6.4 kilowatts. The classes are "good" and "bad" radar returns. "Good" radar returns are those showing evidence of some type of structure in the ionosphere. "Bad" returns are those that do not; their signals pass through the ionosphere. Received signals were processed using an autocorrelation function whose arguments are the time of a pulse and the pulse number. There were 17 pulse numbers for the Goose Bay system. Instances in this database are described by 2 attributes per pulse number, corresponding to the complex values returned by the function resulting from the complex electromagnetic signal. There are 126 instances for "good" class and 225 instances for "bad" class.

This data set is for the diagnosis of the family of Erythemato-squamous diseases which pose a serious problem in dermatology [41]. They all share the clinical features of Erythema and scaling, with very little differences. The diseases in this group are Psoriasis, Seboreic Dermatitis, Lichen Planus, Pityriasis Rosea, Cronic Dermatitis, and Pityriasis Rubra Pilaris.

The Handwritten numerals data set consists of a set of handwritten numerals as used on the Dutch utility maps. They were scanned in 8 bits using 400 dpi. The grey value images were sharpened and normalized for a size resulting in 30×48 binary pixels. The 30×48-pixel tile was divided into 240 (15×16) tiles of 2×3 pixels. All these tiles were averaged, resulting in 240 features. For each of the 10 classes, namely represented by single decimal digits '0' through '9', 200 instances are available.

The Isolet data set contains 7797 instances of spoken letters. The dataset was recorded from 150 speakers balanced for gender and representing many different accents and dialects. Each speaker spoke each of the 26 letters twice (except for a few cases). A total of 617 features were computed for each utterance. Spectral coefficients account for 352 of the features. The features include spectral coefficients; contour features pre-sonorant features, and post-sonorant features [42].

Gisette dataset is from the handwritten digit recognition problem domain. The problem is to separate the highly confusable digits '4' and '9' [41]. The digits have been size-normalized to a fixed-size image of the dimensions 28×28. Pixels of the original data were sampled at random in the middle top part where containing the information necessary to disambiguate 4 from 9. Higher order features were created as products of these pixels to project the problem into a higher dimensional feature space. The data set contains 13500 instances and 5000 attributes. There are reportedly 2500 probe attributes, which have no predictive power.

**Data Preprocessing**

Preprocessing the data was done to improve the efficiency of neural network training [43,44]. Classes in data sets were represented in distributed binary format: for instance, class 3 is represented by the binary sequence "001". Patterns in the training data set were randomly selected for presentation to the neural network to improve the performance.

Normalization is a "scaling down" transformation of the features. Input data normalization prior to training process is crucial to obtaining good results [45]. Within a feature, there is often a large difference between the maximum and minimum values, e.g. 0.01 and 1000. When normalization is performed the value magnitudes are scaled to appreciably small values. In our simulation study, we use the min-max normalization for its lower computational cost [46]. Assuming that for a particular feature $x$, $x'$ is the value after normalization, computation for min-max normalization is shown in Equation 8 where min($x$) is the minimum value of $x$ and max($x$) is the maximum value:

$$x' = \frac{x - \min(x)}{\max(x) - \min(x)} \tag{8}$$

Consequently, attributes or features for every data set are normalized to the range of [-1, 1].

It has been observed that class imbalance (that is, significant differences in class prior probabilities) may result in substantial deterioration of the performance achieved by existing learning and classification algorithms [46.47]. Among the datasets presented in Table 2, the classes for the Dermatology data set are unbalanced. A common way to deal with imbalance is resampling. We employed a technique called SMOTE proposed by the Chalwa et al. [48]. The SMOTE method generates new synthetic data along the lines between the minority examples and their selected nearest neighbors. The original and balanced class distributions for the Dermatology dataset are presented in Figure 8.

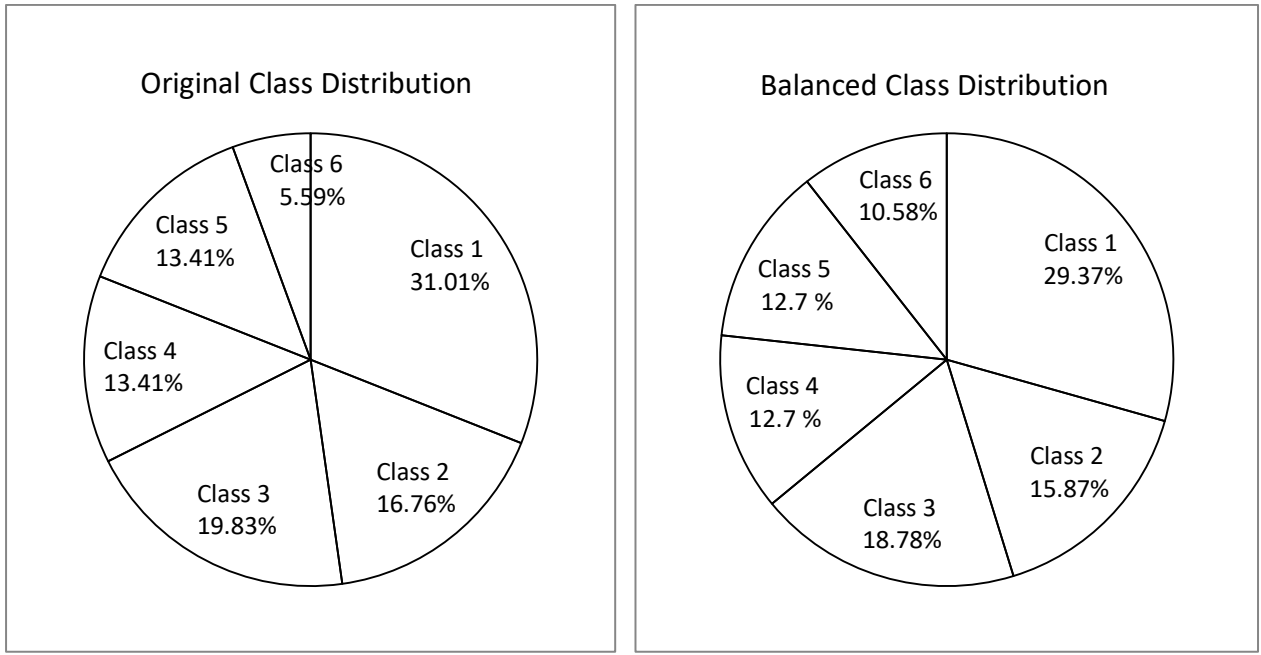

Figure 8. Dermatology data set (a) Original proportion of classes (b) SMOTE balanced classes

We split the data sets into two subsets as the training and the testing data. The training data subset is used for training the network, while the testing data subset is used for testing the performance of the trained network. Two-thirds or 66.67% of the instances are selected as the training data, and one-third or 33.33% as the testing data. The data sets are split according to the proportions of each class in order to make the testing results impartial.

**MLP Neural Network Design**

It is a well-established fact in the literature that the basic back-propagation algorithm for an MLP NN can be very slow for training even a simple multilayer perceptron neural network [37,38]. What is needed is a fast learning algorithm, which projects minimal computational cost, requires minimal centralized coordination, if any; is realizable through incremental learning (vs. batch learning); and can be implemented in a parallel and distributed manner.

The backpropagation with momentum training algorithm responds to all of the requirements listed above. Since the steepest descent is particularly slow when there is a long and narrow valley in the error function surface [39], the momentum term allows a network to respond not only to the local gradient, but also to recent trends in the error surface. The momentum term helps average out the oscillation along the short axis while at the same time adds up contributions along the long axis [40]. It is well known that such a term greatly improves the speed of learning. Momentum is added to back propagation learning by making weight changes equal to the sum of a fraction of the most recent weight change and the new change suggested by the gradient descent rule. When using the back propagation with momentum, the *i*-th update value for a weight $w_i$ is given by

$$\Delta w_i(i) = -\gamma g_i + \alpha \Delta w_i(k-1) \tag{9}$$

where $\gamma$ is the learning rate, $\alpha$ is the momentum rate, $g_i$ is gradient of the error with respect to the weight vector, $\Delta w_i(k-1)$ is the prior update value for the $w_i$, and k is the iteration index. During training of each neuron, all that is needed is to store the previous local weight update values. This makes the training algorithm feasible to implement for distributed and local computation on a WSN.

The ideal context within which the MLP NN would be deployed is one where the tasks need to be accomplished autonomously to the extent possible. Accordingly, the WSN will need to be provided guidelines, heuristics, rules of thumb, bounds or formulas to help aid in the process of initializing or setting its parameter values including the learning rate, momentum and the hidden layer neuron count among others. The primary focus of our study is to demonstrate the feasibility of MLP NN training in a fully distributed and parallel framework on a WSN. As such, establishing specific values for the learning rate, momentum and the hidden layer neuron count that would be applicable for a wide variety of problem domains is of interest. Therefore, following the lead by Weka Machine Learning workbench, we adopted the default settings as suggested by the Weka for the parameters mentioned. We set the learning rate and momentum rate to be 0.3 and 0.8 respectively, same as the default settings of Weka for the MLP classifier [49]. The only exception to these settings are that the learning rate is set as 0.03 for the Isolet and Gisette datasets, which is based on the fact that Weka default settings did not lead to successful training results for our in-house simulations.

For an MLP NN, it is easy to set the input layer and output layer neuron counts since they are the same as the number of attributes in pattern vectors and classes of the data set, respectively. The hidden layer plays a vital role in the performance of MLP NN: it can directly affect the learning and convergence processes. Deciding the number of neurons in the hidden layer is a very important part for setting up an MLP NN [50]. Many researchers invested heavily in finding a well-defined procedure or criterion to decide this number (of hidden layer neurons) over the past several decades. Although there has been some progress, it is not possible to state that one can formulate a number for the hidden layer neuron count readily without a need for empirical exploration first. In the current literature there are empirical formulas [51-54] which suggest reasonably good heuristics to determine the bounds for the number of hidden layer neurons. There is also evidence [50,54] that shows that the number of hidden layer neurons should be determined by training several networks and estimating the generalization error of each. In the case of implementing an MLP NN on a WSN, determining the hidden layer neurons by trial-and-error is not feasible since it would be too costly or unpredictable. It is desirable however to provide at least certain guidelines to make this search or task easier. A formula that sets a range of values on the hidden layer neuron count such that the MLP NN delivers good (but not necessarily superb) performance is the goal.

The current literature proposes several bounds on the hidden layer neuron count. Boger [51] proposed the following formula for the number of hidden layer neurons

$$n_{hid} = \frac{2}{3}(n_{in} + n_{out}), \qquad (10)$$

where $n_{hid}$ denotes the hidden layer neuron count, $n_{in}$ represents the input layer neuron count, and $n_{out}$ is for the output neuron count. According to the Kolmogorov theorem [52], the number of hidden neurons should be

$$n_{hid} = 2 \times n_{in} + 1 \qquad (11)$$

Daqi and Shouyi [53] determined the "best" number of hidden neurons as

$$n_{hid} = \sqrt{n_{in} \times (n_{out} + 2)} \qquad (12)$$

The default setting for the number of hidden neurons in Weka [76] is given by

$$n_{hid} = (n_{in} + n_{out})/2 \qquad (13)$$

In order to determine which formula lead to better performance for our datasets, we conducted an exploratory simulation study to compare the performance of MLP generated by each formula on the first six data sets presented in Table 3. We did not conduct simulation on the Gisette data set since the time cost is prohibitively high for multiple simulations on that data set. For further comparison, we also adopted hidden neuron counts for the same data sets as reported in literature [55-60]. During the experiment, each MLP NN instance is trained with the training data set and evaluated by the test data set. The training stops when the test (validation) data error begins to increase. The training for each MLP was repeated several times with different initial weights. We compared the mean squared error (MSE) values due to the testing data and the iteration count it took to train the network. For Numeral and Isolet datasets, we repeated the training for each MLP instance for 5 times rather than 10 due to the potentially very long training times for these two large data sets. Another point of exception is that the MLP NN generated by the Equation 11 for the Isolet dataset would be too large, which would cost too much time to train. In conjunction with the poor performance of the MLP NN instance generated through Equation 11 for other data sets, it is reasonable to conclude that hidden layer neuron count through this equation would not be suitable for the Isolet dataset either.

In general, MLP NN instances generated by Equation 12, which prescribes a consistently smaller number of hidden layer neurons, always delivered acceptable performance while being remarkably good for the large data sets. Consequently, we chose to use Equation 12 as the formula to determine the number of hidden layer neurons for the simulation study.

**Simulation Results**

In this section, the classification accuracy performance of the WSN-MLP on all seven datasets are compared to the performances of various algorithms as reported in the machine learning literature on the same datasets [61-76]. The purpose of this comparison is to evaluate the performance of WSN-MLP within the larger context of machine learning approaches. In Tables 3 through 9, the best (max) and the worst (min) performance of WSN-MLP as well as the performance of the non-distributed MLP (through in-house implementation) are reported for comparison. These tables show the classification accuracy values ordered in a descending manner, where the values for WSN-MLP and MLP (in-house) are in boldface. Results across all the tables show that the WSN-MLP has a competitive performance with a very diverse and comprehensive set of machine learning classifiers. The maximum performance for the WSN-MLP is among the upper-middle tier of the entire group of machine learning classifiers. It is also worth noting that the WSN-MLP performance even surpasses those of all machine learning classification algorithms for the Dermatology dataset.

Table 3. Comparison of Classification Accuracy for Iris Data Set

| Algorithm | Name | Reference | Accuracy |
|---|---|---|---|
| **QNN** | Quantum Neural Network | [61] | 98.0% |
| **C&S SVM** | Crammer and Singer Support Vector Machine | [62] | 97.3% |
| **SVM** | Support Vector Machine | [63] | 96.7% |
| **WSN-MLP (max)** | | | **96.0%** |
| **MLP** | In-house | | **96.0%** |
| **C4.5** | C4.5 Tree | [63] | 96.0% |
| **NBC** | Naive Bayes Classifier | [63] | 94.0% |
| **LBR** | Logitboost Bayes Classifier | [64] | 93.2% |
| **WSN-MLP (min)** | | | **88.0%** |

Table 4. Comparison of Classification Accuracy for Wine Data Set

| Algorithm | Name | Reference | Accuracy |
|---|---|---|---|
| **SVM** | Support Vector Machine | [62] | 99.4% |
| **LBR** | Logitboost Bayes Classifier | [64] | 98.7% |
| **WSN-MLP (max)** | | | **98.3%** |
| **MLP** | In-house | | **98.3%** |
| **Bayesian Network** | Bayesian Network | [64] | 98.2% |
| **WSN-MLP (min)** | | | **96.7%** |
| **M-SVM** | Multiclass Support Vector Machine | [65] | 96.6% |
| **C4.5** | C4.5 Tree | [66] | 92.8% |
| **M-RLP** | Multicategory Robust Linear Programming | [65] | 91.0% |

Table 5. Comparison of Classification Accuracy for Ionosphere Data Set:

| Algorithm | Name | Reference | Accuracy |
|---|---|---|---|
| **SVM** | Support Vector Machine | [67] | 95.2% |
| **MLP** | In-house | | **94.0%** |
| **Refined GP** | Refined Genetic Programming Evolved Tree | [68] | 92.3% |
| **C4.5** | C4.5 Tree | [68] | 91.1% |
| **WSN-MLP (max)** | | | **90.6%** |
| **Bayesian network** | Bayesian Network | [64] | 89.5% |
| **FSS** | Forward Sequential Selection | [69] | 87.5% |
| **WSN-MLP (min)** | | | **80.3%** |

Table 6. Comparison of Classification Accuracy for Dermatology Data Set

| Algorithm | Name | Reference | Accuracy |
|---|---|---|---|
| **WSN-MLP (max)** | | | **95.3%** |
| **MLP** | In-house | | **95.3%** |
| C4.5+GA | C4.5 Tree with Genetic Algorithm | [70] | 94.5% |
| VFI5 | Voting Feature Intervals | [71] | 93.2% |
| FSS | Forward Sequential Selection | [69] | 90.4% |
| **WSN-MLP (min)** | | | **89.8%** |
| C4.5 | C4.5 Tree | [69] | 86.0% |

Table 7. Comparison of Classification Accuracy for Numerical Data Set

| Algorithm | Name | Reference | Accuracy |
|---|---|---|---|
| **MLP with feature selection method** | MLP with feature selection method | [71] | 98.5% |
| **MLP** | In-house | | **96.6%** |
| **WSN-MLP (max)** | | | **96.4%** |
| KNN | K-Nearest Neighbor | [71] | 95.8% |
| LVQ | Learning Vector Quantization | [72] | 94.9% |
| **WSN-MLP (min)** | | | **94.5%** |
| kNN(PCA) | K-Nearest Neighbor with Principal component analysis | [73] | 80.4% |
| kNN(LDA) | K-Nearest Neighbor with Linear discriminant analysis | [73] | 78.9% |

Table 8. Comparison of Classification Accuracy for Isolet Data Set

| Algorithm | Name | Reference | Accuracy |
|---|---|---|---|
| SVM | Support Vector Machine | [74] | 97.0% |
| kLOGREG | Kernelized logistic regression | [74] | 97.0% |
| **WSN-MLP (max)** | | | **95.9%** |
| **MLP** | In-house | | **95.8%** |
| **WSN-MLP (min)** | | | **94.9%** |
| NBC | Naive Bayes | [75] | 84.4% |
| C4.5 | C4.5 Tree | [75] | 80.2% |
| kNN(LDA) | K-Nearest Neighbor with Principal component analysis | [73] | 71.2% |
| kNN(PCA) | K-Nearest Neighbor with Linear discriminant analysis | [73] | 59.9% |

Table 9. Comparison of Classification Accuracy for Gisette Data Set

| Algorithm | Name | Reference | Accuracy |
|---|---|---|---|
| **LeNet** | Convolutional Neural Network | [76] | 99.2% |
| **MLP (deskewing)** | Deskewing Multilayer Perceptron | [76] | 98.4% |
| **MLP** | In-house | | **97.0%** |
| **WSN-MLP (max)** | | | **96.7%** |
| **MLP** | Multilayer Perceptron | [76] | 96.4% |
| **kNN** | K-Nearest Neighbor | [76] | 95.0% |
| **WSN-MLP (min)** | | | **94.2%** |
| **Linear Classifier** | Logitboost Bayes Classifier | [76] | 88.0% |

**Complexity Analysis for Computation and Communication**

This section will present space, time and messaging complexit analysis for the WSN-MLP design [78]. Space complexity of the WSN-MLP design is minimal which is possible due to distributed implementation of the MLP NN across the motes of a WSN. Each mote allocates adequate memory storage for the requirements of no more than several neurons but as few as one neuron. For maximum parallelism and distributed computation, a single mote would house one neuron. A mote with a single embedded neuron needs to store inputs, which are outputs from all the other neurons, weights and training parameters, and computation model of the neuron. The most costly data structure is the neuron input array, which will grow linearly in the number of neurons. This array is used to store past outputs of all other neurons, which connect to or communicate with a given neuron. Therefore, the space complexity for one-neuron-per-mote case is linear in the number of neurons. If there are multiple neurons housed by a single mote, then the space complexity is still linear.

Time complexity of the WSN-MLP design will be affected by a number of factors including the data set attributes including its size, the classification or approximation function characteristics, training algorithm for the MLP NN among others. Consider a data set with $|P_T|$ patterns in the training set and $|P_V|$ patterns in the validation or testing set. Without loss of generality, the MLP NN will be assumed to have one hidden layer. MLP network learns or adapts its weights after each pattern presentation during training. Due to the WSN context, processing of each pattern is realized in parallel at the level of individual neuron through distributed (and asynchronous) computation. Processing time by individual neurons can be incorporated into the cumulative delay, $t_{wait}$, which is mainly affected by the delays originating due to medium access control and routing protocol implementations. This delay parameter is a random variable and its value depends on numerous factors. Another random variable is the number of iterations, $N_{iter}$, needed for convergence to a solution. Its value depends on the training algorithm, the initial weight and parameter values, error function and the stopping criterion, the data set characteristics and its presentation order among to list the most influential ones. Hence, the time complexity, $C_T$, for a WSN-MLP architecture can be approximated by the following equation:

$$C_T = N_{iter} \times (|P_T| + |P_V|) \times \mathrm{E}\{t_{wait}\}, \qquad (14)$$

where E{} is the expected value operator and $t_{wait}$ parameter value is set in the simulation according to the Equation 6. The mean value of truncated Gaussian distribution, namely $\mu$, is relocated to be 1.0 in the simulation. To get the actual value for time, $\mu$ value needs to be multiplied by the value of per hop delay. From the literature survey [7-28], the range of per hop delay is 2 ms to 226 ms and the expected value is 65 ms. Mean values for the iteration count and the simulation duration were employed to calculate the simulation duration or time complexity values presented in Table 10.

Table 10. Time and message complexity for WSN-MLP design

| Data set | Iteration count | Training pattern count | Testing pattern count | $M_{FP}$ | $t_{wait}$ (ms) | Simulation duration (hrs) | Message packet count |
|---|---|---|---|---|---|---|---|
| Iris | 179 | 100 | 50 | 22 | 187.2 | 1.47 | 985,671 |
| Wine | 194 | 117 | 58 | 52 | 234.0 | 1.73 | 2,858,248 |
| Ionosphere | 126 | 234 | 117 | 45 | 218.4 | 2.28 | 3,260,205 |
| Dermatology | 139 | 239 | 119 | 269 | 312.0 | 4.07 | 23,755,353 |
| Numerical | 113 | 2000 | 666 | 2128 | 468.0 | 28.95 | 804,279,201 |
| Isolet | 170 | 5198 | 2599 | 20724 | 702.0 | 261.90 | 2,549,444,060 |
| Gisette | 106 | 4667 | 2333 | 1446 | 577.2 | 119.88 | 1,381,217,200 |

As Table 10 indicates, among the parameters that appear in the time complexity formula in Equation 14, the number of patterns and consecutively the number of hidden-layer neurons vary significantly. The change in the value of hidden-layer neuron count will result in a change in the value of the parameter $L_{max}$ which in turn results the value of $t_{wait}$ to increase. In overall, simulation duration (or the time cost/complexity) increases linearly with the increase in the pattern count (or neuron count in both hidden and output layers). Therefore, empirical findings are in agreement with the theoretical approximation presented in Equation 14.

Message complexity associated with the neurocomputing is measured by estimating the number of messages sent which carry neuron output values since this aspect contributes as the major component of the communication cost. More specifically, each original or retransmission of a given message that carries a neuron output value is counted as a basic unit of measurement. During forward propagation cycle of the MLP NN training phase through backpropagation with momentum, $N_{hid}$ neurons in the hidden layer send their outputs on the average to $N_{out}$ neurons in the output layer. Each wireless communication handover or hop requires retransmission of a given message. The hop distance between hidden neuron $i$ and output neuron $j$ will be denoted by $h_{ij}$. Therefore, the total number of message transmissions (including retransmissions) for a single training pattern during forward propagation cycle as represented by $M_{FP}$ is given by

$$M_{FP} = \sum_{i=1}^{N_{hid}} \sum_{j=1}^{N_{out}} h_{ij}.$$

This cost is incurred for each training pattern in the training and validation sets, namely $P_T$ and $P_V$. The message complexity, $C_M$, for the entire training episode will depend on the size of the data set as well as problem attributes, which, in conjunction with other factors such as the initial values of weights and learning parameters to name a few, will dictate the number of iterations to convergence. Accordingly, the message complexity for the forward propagation phase is estimated by the following equation:

$$C_{M,FP} = N_{iter} \times (|P_T| + |P_V|) \times M_{FP}.$$

During the backward propagation phase, $N_{out}$ output-layer neurons transmit their outputs to $N_{hid}$ hidden-layer neurons, which results in the same number of messages as $M_{FP}$. Assuming that online or incremental learning is implemented, the above cost is incurred for each pattern in the training set for the duration of training. Hence, the message complexity for the backward propagation phase is estimated by

$$C_{M,BP} = N_{iter} \times |P_T| \times M_{FP}.$$

The overall message complexity for the training and validation combined is given by

$$C_M = C_{M,FP} + C_{M,BP} = N_{iter} \times (2|P_T| + |P_V|) \times \sum_{i=1}^{N_{hid}} \sum_{j=1}^{N_{out}} h_{ij}. \qquad (15)$$

As the Equation 15 indicates, the message complexity depends on number of iterations, the number of training and testing patterns, and the sum over the hop counts between every node pair.

The messaging measurements for all data sets are presented in Table 10. As the number of neurons (which is calculated as a function of the pattern count) in the hidden and output layers increases, the sum of distances (as measured by the number of hops) between any two neurons (motes) also increases as does the number of messages (packets). For a hundred-fold increase in the neuron count, the total hop count increases thousand-fold, and the number of messages increases approximately thousand-fold as well. The message complexity is worse than linear, and therefore, the message complexity sets the dominant bound for the scalability of the WSN-MLP algorithm compared to the time and space complexities.

## V. Conclusions

This paper presented a methodology for simulation of wireless sensor networks for application-level software development with the goal of minimizing the computational cost of simulation. All lower-level operational aspects of wireless sensor networks were modeled in terms of their effects at the application layer as message delay and drop. Such models were derived from empirical data as reported in the literature by other researchers for a wide variety of wireless sensor network protocols and topologies. Proposed methodology was demonstrated through a case study, which employed a wireless sensor network as a parallel and distributed neurocomputing platform for a multilayer perceptron neural network. This neural network was configured as a classifier and tested on a set of data from the machine learning repository. Results demonstrated that the proposed methodology promotes substantial reduction in the computational cost associated with simulation of wireless sensor networks and led to the training of multilayer perceptron neural network based classifiers with competitive performance and manageable computation cost for the simulation effort.